\documentclass[10pt]{elsart}
\usepackage{amssymb,amsmath,float}
\usepackage{epsfig,graphics,subfigure}

\begin{document}
\begin{frontmatter}

\title{Matter-wave interferometry in periodic and quasi-periodic
arrays}
\author{Y. Eksioglu},
\author{P. Vignolo\corauthref{cor1}} and
\author{M.P. Tosi}
\corauth[cor1]{Corresponding author, e-mail: {\tt vignolo@sns.it}}
\address{NEST-INFM and Scuola Normale Superiore,
Piazza dei Cavalieri 7, I-56126 Pisa, Italy}
\maketitle

\begin{abstract}
We calculate within a Bose-Hubbard tight-binding model the matter-wave 
flow driven by a constant force through a Bose-Einstein condensate 
of $^{87}$Rb atoms in various types of
   quasi-onedimensional arrays of potential wells. Interference patterns 
are obtained when beam splitting is induced by creating energy minigaps 
either through period doubling
or through quasi-periodicity governed by the Fibonacci series.
The generation of such condensate modulations by means of optical-laser
structures is also discussed.
\end{abstract}

\end{frontmatter}
\newpage
\section{Introduction}
A Bose-Einstein condensate (BEC) is a gas in which a macroscopic number of 
massive particles reside in the same quantum state (see~\cite{Minguzzi2004a} 
for a review of work done on quasi-pure BEC's produced since 1995). 
Experiments aimed at revealing the coherence of a BEC have demonstrated 
its matter-wave properties. In particular, condensate
   interferometry can be realized by splitting a BEC into two parts with 
a definite phase relationship, these parts being then brought into overlap 
and interference as for an optical laser beam that has gone through 
a beam splitter. Coherent splitting of a BEC has been achieved by optically 
induced Bragg diffraction~\cite{Kozuma1999a} and a number of ingenious 
methods have been 
devised to extract a collimated beam of atoms from a BEC 
(see {\it e.g.}~\cite{Inouye1999a}).
   
A quasi-onedimensional (1D) array of potential wells is created for 
an atomic
BEC by the interference 
of two optical laser beams which counterpropagate along the $z$ axis, say, 
and are superposed on a highly elongated magnetic trap.
Such an optical lattice provides an 
almost ideal periodic potential and has allowed the study of 
Bloch and Josephson-like
oscillations~\cite{BenDahan1996a,Anderson1998a,Cataliotti2001a} 
and of the mechanisms by which decoherence arises as in the transition from a superfluid to a Mott-insulator state~\cite{Greiner2002a}.
   
The 1D optical lattice can be modified by means of auxiliary 
laser beams~\cite{Vignolo2003a}. In particular, its periodicity can 
be doubled by adding two beams that are rotated by angles of
 $60^{\circ}$ and $120^{\circ}$ with respect to the $z$ axis. 
For a suitable choice of the phases the potential 
seen by the BEC atoms takes the shape
\begin{equation}
U(z)=U_0[\sin^2(kz)+\beta^2\sin^2(kz/2)]\,,
\label{doublepot}
\end{equation}
where $U_0$ is the potential well depth, $\beta^2$
is the relative energy difference between adjacent wells, 
and $k$ is the laser wavenumber determining the distance $d$ of 
adjacent wells as $d =\pi/k$.
In solid-state terminology, the doubling of the period when $\beta^2\neq0$
causes the 
opening of a minigap in the energy spectrum as a function of $\beta^2$.
A BEC driven through such a lattice by a
constant force is coherently split by a combination of Bragg diffraction 
and of tunnelling through the minigap. In steady state a 
prominent interference pattern is produced and can be observed
by monitoring the outgoing particle flow~\cite{Vignolo2003a}.
   
In this work we show how a quasi-periodic Fibonacci array of potential 
wells could be created by optical means and evaluate its 
density-of-states structure to display a series of approximate minigaps. 
We then show that this structure, unlike a simple periodic structure 
but similarly 
to a period-doubled structure, leads to an interference
pattern under steady-state drive of the BEC by a constant force. 
The model and the behaviours of periodic arrays are briefly reviewed in 
Sec. 2 and Sec. 3, respectively. The
Fibonacci array is treated in Sec. 4, while Sec. 5 offers some concluding 
remarks.

\section{The model}
We use a 1D tight-binding Hamiltonian for the BEC atoms and a 
Green's function approach to evaluate their linear transport 
coefficient through the array of potential wells~\cite{Vignolo2003a}. 
The Bose-Hubbard Hamiltonian for $N$ bosons distributed inside $n_s$ wells is
 \begin{equation}
H_I=\sum_{i=1}^{n_s} 
\left[E_i |\,i\rangle\langle i\,|+\gamma_i(|\,i\rangle
\langle i+1\,|+|\,i+1\rangle
\langle i\,|)\right]\,.
\label{Hamiltonian}
\end{equation} 
Here, the parameters $E_i$ and $\gamma_i$ represent site energies and 
hopping energies, respectively, and depend on the number of bosons in the well labelled by the index $i$. In a
   tight-binding scheme the 1D condensate wavefunction in the $i$-th well 
is a Wannier function for the bosons in the external potential and, 
according to the early work of Slater~\cite{Slater1952a}, can be 
written as a Gaussian function of longitudinal width $\sigma_{z}$,
\begin{equation}
\phi_i(z)=\phi_i(z_i)\exp[-(z-z_i)^2/(2\sigma_{z}^2)]\,.
\label{Wannier}
\end{equation}
The parameters entering the effective Hamiltonian are given by
 \begin{equation}
\hspace{-2.5cm}E_i=\int dz\,\phi_i(z)\left[
-\frac{\hbar^2\nabla^2}{2m}+U(z)+\frac{1}{2}g_{bb}
|\phi_i(z)|^2-ma z+{\mathcal C}\right]\phi_i(z)
\label{siteenergy}
\end{equation} 
   for the site energies and by
\begin{equation}
\hspace{-2cm}\gamma_i=\int dz\,\phi_i(z)
\left[
-\frac{\hbar^2\nabla^2}{2m}+U(z)+\frac{1}{2}g_{bb}
|\phi_i(z)|^2+{\mathcal C}\right]
\phi_{i+1}(z).
\label{hopenergy}
\end{equation} 
   for the hopping energies. 
In Eqs.~(\ref{siteenergy}) and (\ref{hopenergy}) $m$ is the boson mass,
$a=F/m$ is the acceleration due to a constant external force $F$
acting on the bosons, $g_{bb}$ is the effective 1D
   boson-boson interaction parameter, and ${\mathcal C}$
is a constant accounting for transverse effects in a cigar-shaped 
trap (for the determination of the parameters see Ref.~\cite{Vignolo2003a}).
We remark that in a tight-binding approach nonlinear
interaction effects enter the self-consistent calculation of the axial 
width $\sigma_z$, resulting in a broadening of the Gaussian function, 
and also modify the on-site energies and the
hopping energies. This approach is justified in the case of 
weak boson-boson coupling as for a $^{87}$Rb BEC, on which we 
focus in this paper, and should be improved in a
strong-coupling situation as is met on the approach to a Feshbach 
resonance~\cite{Trombettoni2001a}.
   
In the Green's function method the calculation of the transmittivity 
of bosonic matter waves through the array of potential wells does not require 
an explicit solution of the
Hamiltonian (\ref{Hamiltonian}). The array is reduced by 
a renormalization/decimation technique to a single 
``dimer"\cite{Farchioni1996a}, to which an incoming lead and an outgoing 
lead are connected.
The steady-state transport coefficient is inferred from the scattered 
wavefunction of the leads in the presence of the dimer.

Period-doubling of the array as described by the expression of 
$U(z)$ in 
Eq. (\ref{doublepot}) yields an interference pattern as a function of 
the ratio $T_B/\tau$, where $T_B=\hbar k/ma$ is the
period of Bloch oscillations and $\tau$ is the average time needed for 
tunnelling twice across the minigap. The array acts in this case as 
an interferometer for bosonic matter
waves and we construct below its optical analog. 
In essence the minigap plays the role of a medium with large refractive 
index, across which an evanescent wave couples
two layers that allow real wave propagation. In the case of a Fibonacci 
array, on the other hand, periodicity is lost but quasi-periodicity 
induces the opening of a number of rather sharp
depressions in the density of states (``quasi-minigaps"). 
The resulting fragmentation of the spectral density modifies the 
interference pattern, but does not erase it.

\section{Interference from period doubling}
\begin{figure}[H]
\centering{
\epsfig{file=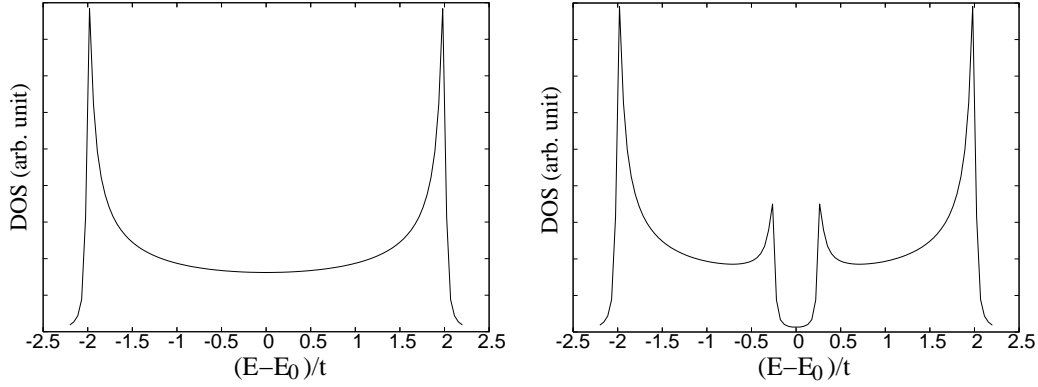,width=1.\linewidth}}
\caption{Total DOS of a very long lattice
(1000 sites) with a single period (left panel) and a doubled period
(right panel), as a function of energy $E$
referred to the central energy $E_0$ and with $4t$ being the
total spectral width. In the single-period lattice $E_0$ is the site energy
$E_i$ and $t$ the hopping energy $\gamma_i$, while 
in the double-period lattice $E_0=(E_i+E_{i+1})/2$
and $t=[(E_i-E_{i+1})^2/2+4\gamma_i^2]^{1/2}/2$.}
\label{fig4}
\end{figure}

When $\beta^2=0$ the ideal infinite lattice generated by $U(z)$
has a single period and its low-energy
spectrum is that of a one-band system,
as is shown in the left panel of Fig.~\ref{fig4}.
Period doubling causes the opening of a minigap
in the total density of states (DOS) (see right panel in Fig.~\ref{fig4}).
The calculation of the DOS
has been carried out by recursive algorithms
as those in Refs.~\cite{Vignolo1999a,Farchioni2000b} and full details 
will be given in a later publication.
The energy width $\Delta E$ of the minigap is fixed by the energy
difference 
$|E_i-E_{i+1}|$ between two adjacent sites and in the limit of an infinite
lattice no states are present inside the minigap.
In real systems as in the experiments at LENS
(see for example ~\cite{Fort2003a})
the condensate occupies about 100-200 wells. As a result of finite-size
effects the gap is not completely empty and the bosons can easily be
transferred by tunnelling between the two sub-bands. 

\begin{figure}[H]
\centering{
\epsfig{file=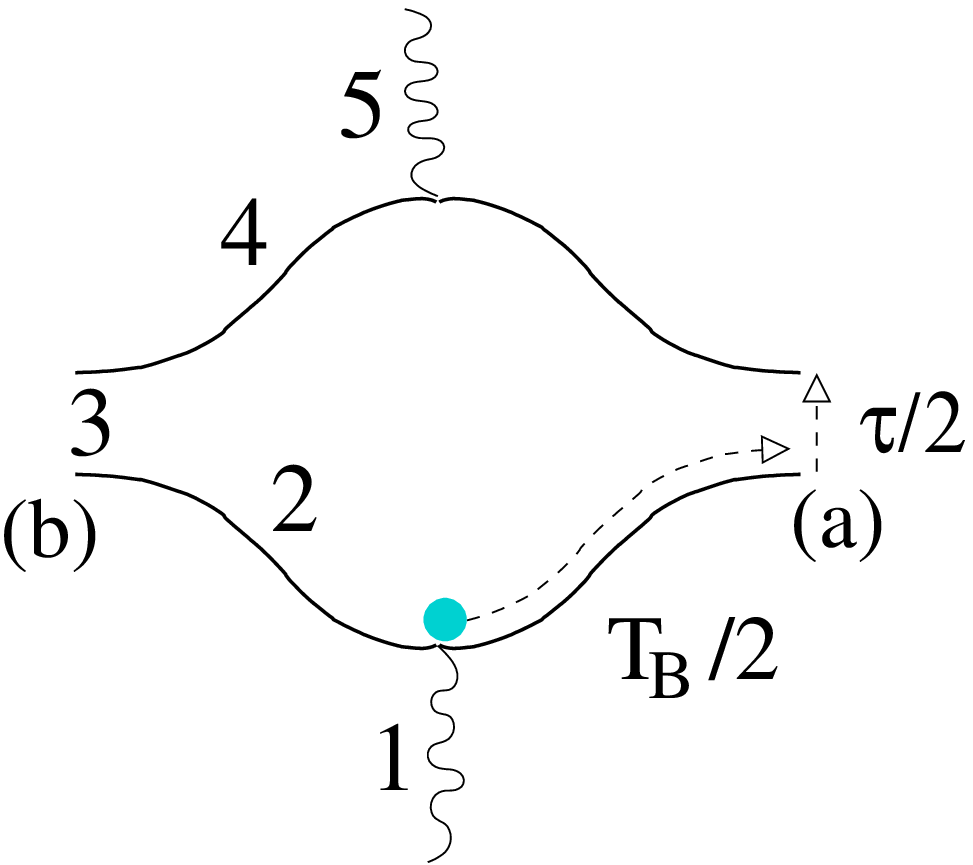,width=0.3\linewidth}
\vspace{0.5cm}

\epsfig{file=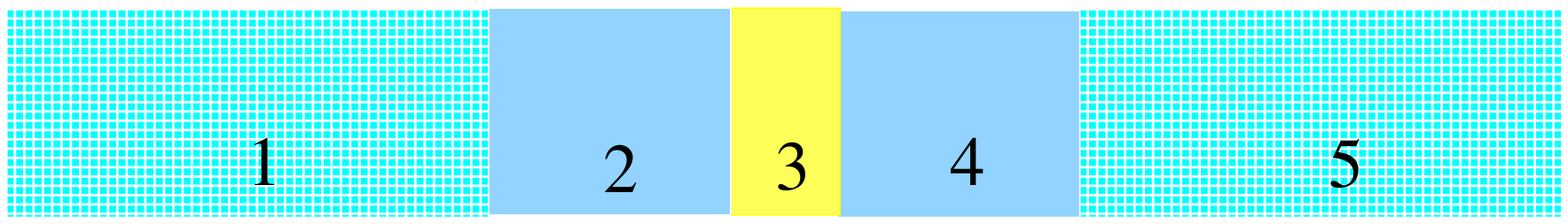,width=0.8\linewidth}}
\caption{Top: schematic representation of the two-band system connected
to an incoming lead (1) and an outgoing lead (5), showing the 
half-period $T_B/2$ of Bloch
oscillations and the tunnelling time $\tau/2$. Bottom: the
equivalent birefringent system for light propagation.}
\label{fig3}
\end{figure}

The addition of a potential $maz$ causes a tilt of the 
bands in space and the density
of states depends on both position and energy.
Thus the condensed bosons are driven through the lattice and explore 
the whole band (in the single-period lattice) or both
sub-bands (in the doubled-period lattice).
On reaching the upper-energy state they are partly allowed to leave 
the lattice towards the continuum. 
For the evaluation of the number of transmitted particles we 
connect the system 
to incoming and outgoing leads, which mimic its coupling 
to the continuum by injecting and extracting a steady-state particle
current. A schematic representation of the two-band situation is given
in the top part of Fig.~\ref{fig3}.

\begin{figure}[H]
\centering{
\epsfig{file=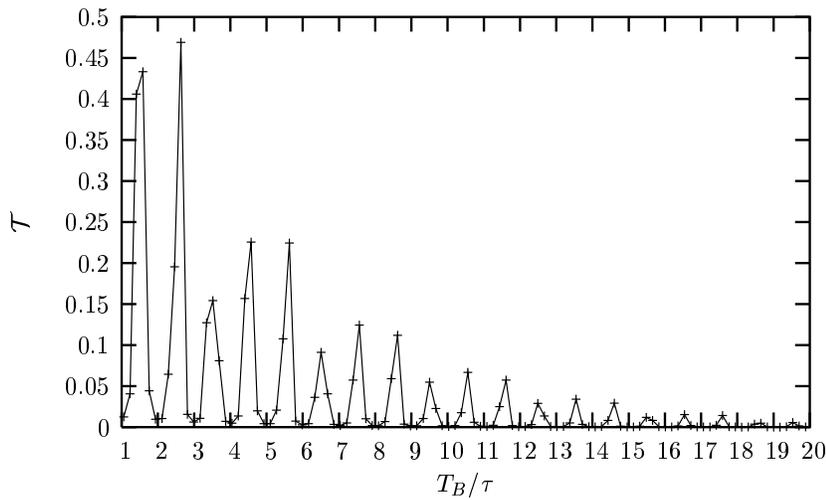,width=0.8\linewidth}}
\caption{Interference pattern in 
condensate transmittivity from period doubling, as a function of
$T_B/\tau$.}
\label{fig6}
\end{figure}

The transmittivity of a BEC of $^{87}$Rb atoms has been evaluated
by using the scattering matrix formalism adapted to the case
of out-of-equilibrium leads~\cite{Vignolo2003a}.
We have used the set of parameters $U_0=3.5\,E_r$ with $E_r=\hbar^2k^2/2m$,
$\beta^2\simeq 0.01$ and $k=8.2\,\mu$m$^{-1}$. 
In the single-period case the particle current varies monotonically
with the external force and hence with the period of Bloch oscillations.
After period doubling the boson wavepacket is split
at the edge of the Brillouin zone (point (a) in Fig.~\ref{fig3}),
where it can either be Bragg-reflected to point (b)
or tunnel into the upper sub-band. The interference
between the wavepackets reaching point (b) by these various paths 
gives the outgoing current
shown in Fig.~\ref{fig6}. The  
minima in transmittivity towards the continuum are located 
at integer values of $T_B/\tau$,
where $\tau=(3\pi^2/8)(\hbar N/n_s\Delta E )$.

\subsection{Optical equivalent}
A condensate wavepacket propagating in an infinite doubled-period lattice 
is equivalent to a light beam of frequency $\omega$
travelling through a five-layer optical medium~(see Fig.~\ref{fig3}).
The first and the last medium in the bottom part of Fig.~\ref{fig3}
 are semi-infinite and play the role of the 
two leads.
The second and the fourth layer stand for the two energy bands, while
the middle layer mimics the minigap.

Let $t_{i,j}$ and $r_{i,j}$ be the transmission and reflection coefficient
at the interface
between the media $i$ and $j$, connected by
$t_{i,j}=1+r_{i,j}$. For a suitable coupling between the leads and
the lattice $t_{1,2}$ and $t_{4,5}$ have unitary modulus
and on crossing these interfaces the wave 
takes up an irrelevant phase factor. 
At the 2-3 and 3-4 interfaces we can mimic the effect
of the minigap in Bragg-reflecting the BEC by imposing total
reflection of light {\it via}
$r_{2,3}=\exp(i\alpha_{2,3})$ and
$r_{3,4}=\exp(i\alpha_{3,4})$, and allow for tunnel by
propagation through layer
3 only {\it via} an evanescent wave.
If the media 2 and 4 have the same refractive index
and the same optical depth, we have $\alpha_{3,4}=\alpha_{2,3}-\pi$
and by symmetry we can set $\alpha_{2,3}=
-\alpha_{3,4}=\pi/2$.
We can then use the recursive 
formula~\cite{Born1959a} 
\begin{equation}
\left\{\begin{array}{l}
r_{i,j+2}=\dfrac{r_{i,j+1}+r_{j+1,j+2}e^{2i\alpha_{j+1}}}
{1+r_{i,j+1}r_{j+1,j+2}e^{2i\alpha_{j+1}}}\\
t_{i,j+2}=\dfrac{t_{i,j+1}+t_{j+1,j+2}e^{i\alpha_{j+1}}}
{1+r_{i,j+1}r_{j+1,j+2}e^{2i\alpha_{j+1}}}\\
\end{array}
\right.
\end{equation}
to calculate the transmission coefficient between the media 1 and 5,
where $\alpha_j$ is the phase shift acquired by light travelling through
the $j$-th layer. To pursue the analogy with the doubled-period lattice
we have set $\alpha_2=\alpha_4=\omega T_B/2$ and 
$\alpha_3=i\omega\tau/2$, with $\tau=2\pi/\omega$.
The total transmission coefficient $|t_{1,5}|^2$ for the light
intensity
is then found to have minima when the ratio $T_B/\tau$ takes integer values,
as in the case of the doubled-period lattice.

The main difference between the two patterns is that in the BEC case
the height of the peaks is largest at low values of  
$T_B/\tau$ (see Fig.~\ref{fig6}), whereas in the optical analog all 
peaks have the same height.
Decreasing $T_B$ leads to a decrease of the optical
depth of media 2 and 4 in the five-layer system, 
while for the condensate it means that the bosons leave the lattice towards
the continuum after having travelled through a lower number of sites.
Therefore, decreasing $T_B$ is equivalent in this case to shortening the array
and hence helps the tunnelling.

\section{Interference from a Fibonacci chain}
\begin{figure}[H]
\centering{
\epsfig{file=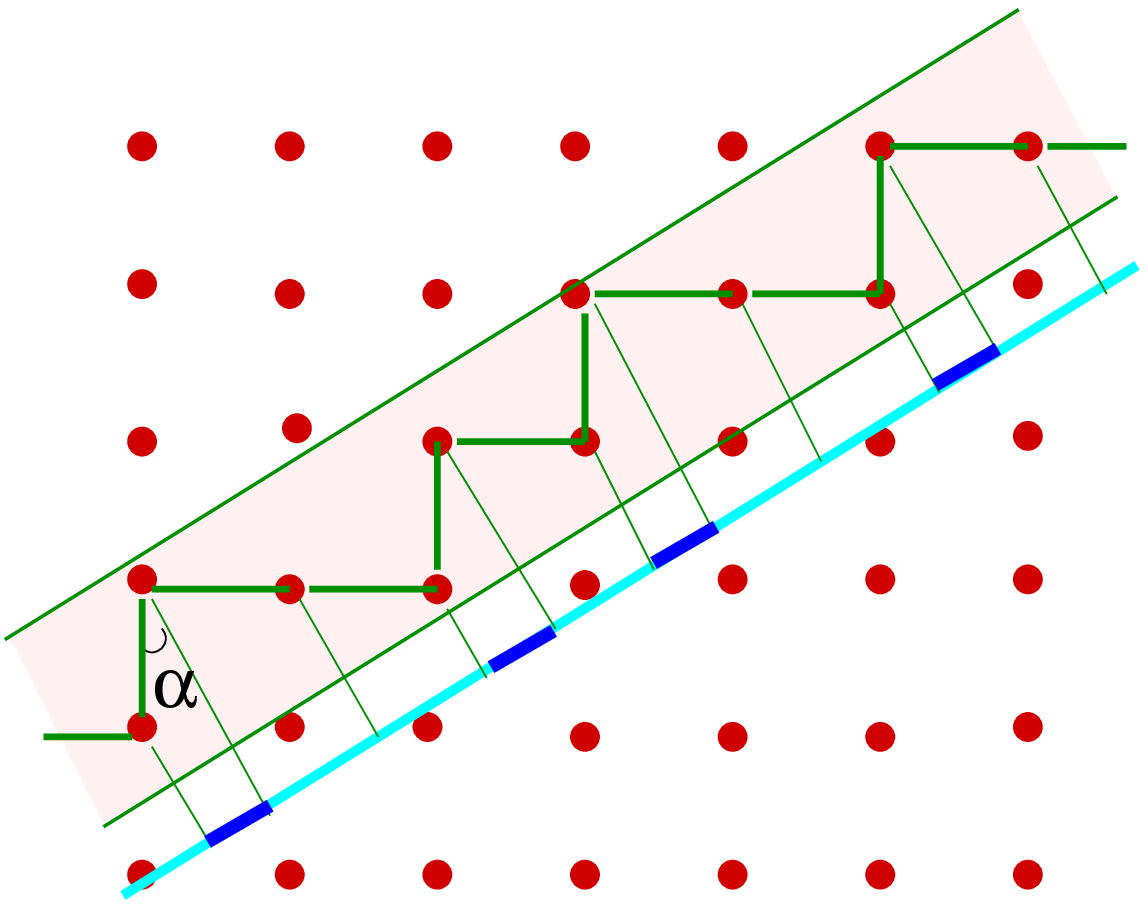,width=0.45\linewidth}
\hspace{1cm}\epsfig{file=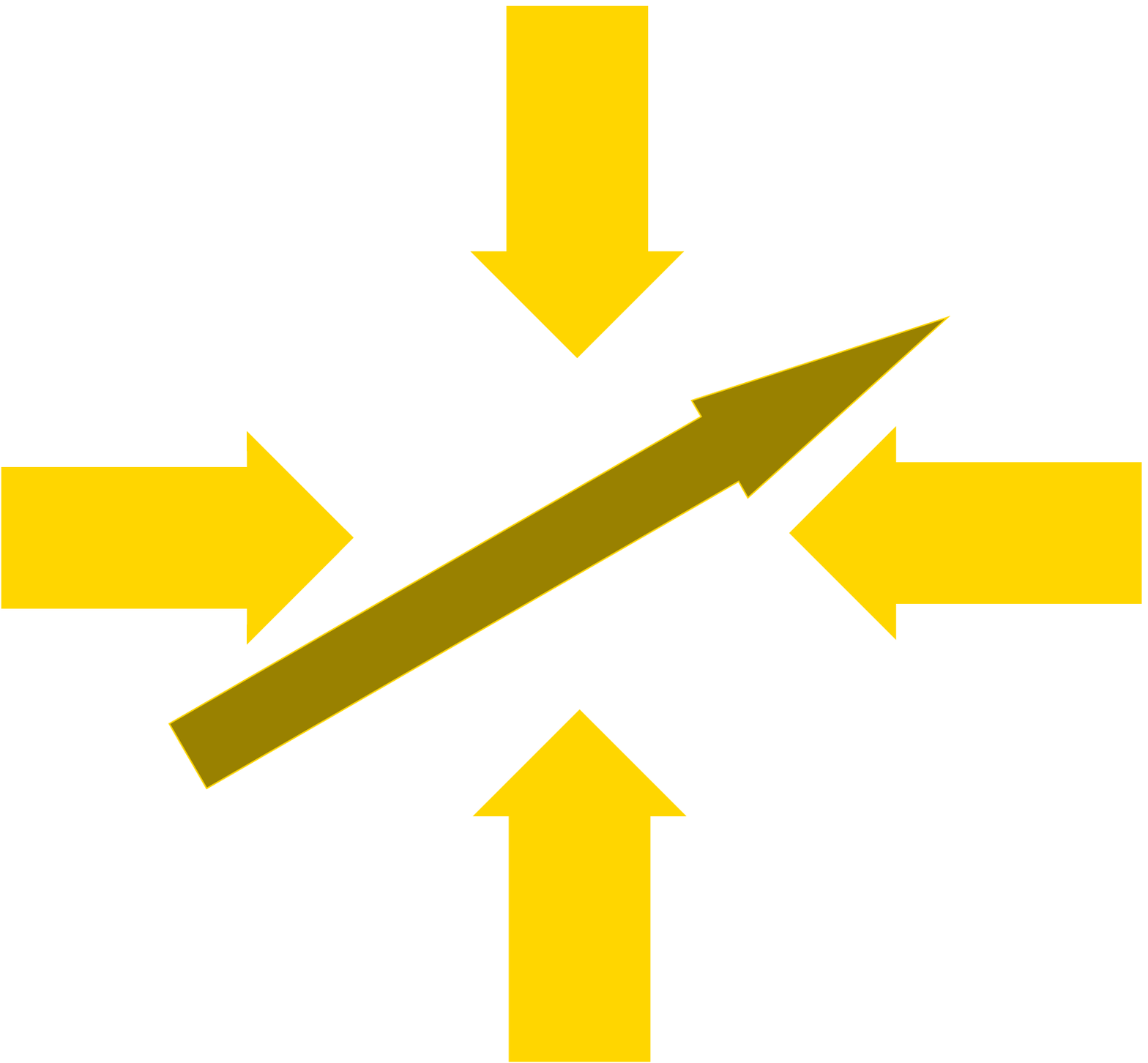,width=0.45\linewidth}}
\caption{Left panel: the projection method for creating the Fibonacci
chain. Right panel: the equivalent set-up
for a Fibonacci optical chain realized within five laser beams.}
\label{fig7}
\end{figure}

The symmetry of an 
optical potential that is created by the interference
of optical laser beams is completely determined by the geometric
arrangement of the beams. Therefore, one can 
not only realize lattices with various symmetries in the laboratory, 
but also design quasiperiodic optical potentials. 
Here we give as an example the use of the projection method 
that takes origin from solid state physics for generating a Fibonacci chain
(see left panel in Fig.~\ref{fig7}).
The 1D Fibonacci arrangement is obtained from a 2D periodic square lattice
by projecting all sites belonging to a strip 
with the irrational
slope ${\alpha={\arctan(2/(\sqrt{5}+1))}}$ onto a line with the same 
slope (see for instance ~\cite{Fujiwara1990a}).
With this construction the distance between two neighbouring projected sites 
can take two different values, $A$ or $B$ say. 
Their ratio $A/B=(\sqrt{5}+1)/2$ is the so-called golden ratio 
and the sequence of distances
follows the Fibonacci chain rule $A B B A B A B B \cdots$.
This sequence can be obtained by the transformation 
rule $A\rightarrow B$ and $B\rightarrow BA$. 

This method can be applied to an atomic gas by using a five-laser 
configuration as is shown in the right panel of Fig.~\ref{fig7}.
Four laser beams build the 2D square lattice by their interference.
The fifth
laser is aligned with the longitudinal axis of the cigar-shaped 
magnetic trap
and drives the condensed bosons along the direction
of the Fibonacci array.
Within this model the hopping energies follow the Fibonacci
sequence, but in order to make a direct comparison with the results on the 
periodic 1D lattices we consider below the case in which
the site energies (rather than the bond lengths) form
the Fibonacci chain. This scheme should at least qualitatively give
the correct physical picture of matter waves propagating through
a quasi-periodic array. 

\begin{figure}[H]
\centering{
\epsfig{file=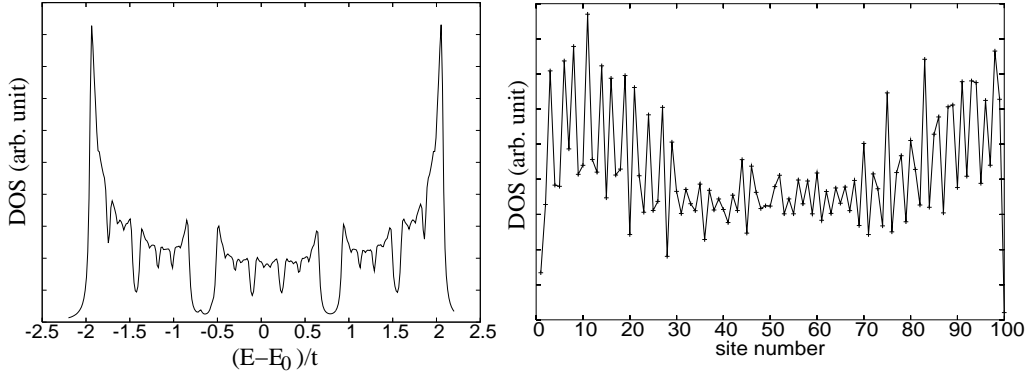,width=1.\linewidth}}
\caption{Left panel: total DOS of a very long
Fibonacci chain (1000 sites) as a function of $(E-E_0)/t$.
Right panel: projected DOS
for a BEC driven by a constant force through a Fibonacci chain of 
100 sites at constant energy $E=E_0-2t$, as a function of the site number.}
\label{fig8}
\end{figure}

The total density of states for a very long Fibonacci 
array is shown in the left panel of Fig.~\ref{fig8}. The fragmentation
of the spectrum is typical of quasi-periodic and aperiodic systems,
as it is well known from previous solid state studies (see for instance
Ref. \cite{Vidal2001a}).
In particular, in the classical case of a quasi-periodic Fibonacci chain 
the spectrum is known to be a Cantor set with measure zero. 
For a chain of 100 wells, the site-projected density of states
(see right panel in Fig.~\ref{fig8}) has a general
envelope resembling that of the single-period lattice,
but is modified by the quasi-periodicity favouring
the population of certain sites.

\begin{figure}[H]
\centering{
\epsfig{file=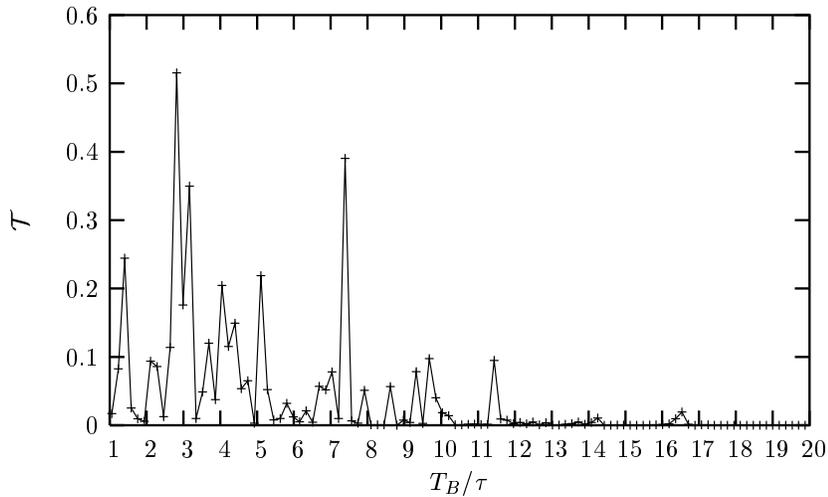,width=0.8\linewidth}}
\caption{Interference pattern in condensate transmittivity through 
a Fibonacci chain, 
as a function of $T_B/\tau$.}
\label{fig9}
\end{figure}

The condensed bosons travelling through the chain explore an energy spectrum
which on average is rather more akin to the single-period band structure
than to the
doubled-period one. Nevertheless, the transmittivity through
a Fibonacci chain presents an interference pattern (see Fig.~\ref{fig9}).
The presence of peaks is a signature of quasi-periodicity leading to
``quasi-minigaps'' in the energy spectrum.

\section{Conclusions}
In summary, we have reviewed earlier work on condensate transport
in optical lattices and proposed an optical equivalent for the
interference pattern that is generated by the opening of a minigap
in the energy spectrum when matter waves are propagating in a
lattice with doubled period. 

We have then shown that condensate
interference also results from the opening of sharp depressions in
the spectral density of states for a matter wave
propagating in a quasi-periodic array.
Although quasi-periodicity is often said to be in some sense intermediate
between perfect periodicity and complete disorder, it has been shown
in earlier work~\cite{Vignolo2003a} that in neither of these
two cases an interference
pattern of any sort is found in the absence of minigaps. 
We have also proposed a method by which
a quasi-periodic modulation of the Fibonacci type may be created by optical
means for a condensate in an elongated magnetic trap.

\ack{This work was partially supported by INFM through the PRA-Photonmatter.}

\end{document}